\documentclass[pra,twocolumn,amsmath,amssymb,superscriptaddress,showpacs]{revtex4}
\usepackage[dvips]{color}
\usepackage{graphicx}
\usepackage{epsfig}
\usepackage{dcolumn}
\usepackage{bm}

\newcommand{\im}{\mathop{\mathrm{Im}}}

\renewcommand{\d}{\mathrm{d}}

\newcommand{\be}{\begin{equation}}
\newcommand{\ee}{\end{equation}}
\newcommand{\bea}{\begin{eqnarray}}
\newcommand{\eea}{\end{eqnarray}}
\newcommand{\bse}{\begin{subequations}}
\newcommand{\ese}{\end{subequations}}

\newcommand{\pf}{k_{\mathrm F}}
\newcommand{\kf}{k_{\mathrm F}}
\newcommand{\ef}{\varepsilon_{\mathrm F}}

\begin{document}

\title{Dynamic and static density-density correlations in the one-dimensional Bose gas:\\
exact results and approximations}

\author{Alexander~Yu.~Cherny}
\affiliation{Bogoliubov Laboratory of Theoretical Physics, Joint Institute for Nuclear
Research, 141980, Dubna, Moscow region, Russia}

\author{Joachim~Brand}
\affiliation{Centre for Theoretical Chemistry and Physics and Institute of Fundamental
Sciences, Massey University, Auckland, New Zealand}

\date{\today}

\begin{abstract}
We discuss approximate formulas for the dynamic structure factor of the one-dimensional 
Bose gas in the Lieb-Liniger model that appear to be applicable over a wide range of the 
relevant parameters such as the interaction strength, frequency, and wavenumber. The 
suggested approximations are consistent with the exact results known in limiting cases. 
In particular, we encompass  exact edge exponents as well as Luttinger liquid and 
perturbation theoretic results. We further discuss derived approximations for the static 
structure factor and the pair distribution function $g(x)$. The approximate expressions 
show excellent agreement with numerical results based on the algebraic Bethe ansatz.
\end{abstract}
\maketitle

\section{Introduction}

Correlations in ultracold atomic gases arise from the interplay of quantum statistics, 
interactions and thermal and quantum fluctuations. Recently, a lot of progress has been 
made experimentally to probe and characterize these correlations 
\cite{stenger99,altman04,bloch:885,gericke08}. The one-dimensional Bose gas is a 
particularly interesting system as quantum correlations generally play a larger role 
compared to three dimensional Bose-Einstein condensates and regimes with very different 
correlation properties can be probed experimentally~\cite{paredes04,weiss04}. In these 
experiments, elongated ``spaghetti'' traps are created by optical lattices, which 
confine the atomic motion in the transverse dimensions to zero-point quantum 
oscillations~\cite{goerlitz01}. Thus, the systems become effectively one-dimensional.

Theoretically, interactions of the rarefied atoms in one-dimensional waveguides are well 
described by effective $\delta$-function interactions \cite{olshanii98}. The resulting 
model of a one-dimensional Bose gas is an archetype of an integrable but non-trivial 
many-body system that has been receiving long standing interest from physicists and 
mathematicians alike. The model was first solved with Bethe ansatz by Lieb and Liniger 
\cite{lieb63:1,lieb63:2}, who calculated the ground-state and excitation energies. 
Depending on the value of the dimensionless coupling strength, the Lieb-Liniger model 
describes various regimes with the corresponding correlations. Being exactly solvable, 
the model, however, does not admit complete analytic solution for the correlation 
functions. Up-to-now, this is complicated and challenging problem in 1D 
physics~\cite{korepin93:book,giamarchi04:book}.

Dynamical density-density correlations can be measured in cold atoms by the two-photon 
Bragg scattering \cite{stenger99,ozeri04}. Theoretically, they are described by the 
dynamic structure factor (DSF) \cite{pitaevskii03:book}
\begin{equation}
S(k,\omega)=L\int \frac{\d t\d x}{2\pi\hbar}\,e^{i(\omega t-k x)}
\langle0|\delta\hat{\rho}(x,t)\delta\hat{\rho}(0,0)|0\rangle , \label{eqn:dsfdef}
\end{equation}
where $\delta\hat{\rho}(x,t)\equiv\hat{\rho}(x,t)-n$ is the operator of density fluctuations 
and  $n=N/L$ is the equilibrium density of particles. We consider zero temperature, 
where  $\langle0|\ldots|0\rangle$  denotes the ground-state expectation value. The DSF 
is proportional to the probability of exciting a collective mode from the ground state 
with the transfer of momentum $k$ and energy $\hbar\omega$, as can be seen from the 
energy representation of Eq.~(\ref{eqn:dsfdef})
\begin{equation}
S(k,\omega)=\sum_m |\langle0|\delta\hat{\rho}_k|n\rangle|^2\delta(\hbar\omega-E_m+E_0),
\label{eqn:dsfenergy}
\end{equation}
where $\delta\hat{\rho}_k=\sum_{j}e^{-i k x_j}-N\Delta(k)$ is the Fourier component of 
$\delta\hat{\rho}(x)$, $\Delta(k)=1$ at $k=0$ and $\Delta(k)=0$ otherwise. Once the DSF 
is known, the static structure factor $S(k)$ and the pair distribution function $g(x)$ 
can be calculated by integration as is discussed in Sec. III.B.

Previously known results for the DSF of the one-dimensional Bose gas come from Luttinger 
liquid theory, which predicts a power-law behavior of the DSF at low energies in the 
vicinity of the momenta $k=0,2\pi n, 4\pi n\ldots$ and yields universal values for the 
exponents \cite{haldane81,castro_neto94,astrakharchik04}. In the regime of strong 
interactions, we have previously derived perturbatively valid expressions  covering 
arbitrary energies and momenta at zero \cite{brand05} and finite 
temperature~\cite{cherny06}. For finite systems, it is possible to compute the 
correlation functions numerically, using the results of the algebraic Bethe ansatz 
calculations \cite{caux06,caux07}. Finally, the exact power-law behavior along the 
limiting dispersion curve of the collective modes has recently been calculated in 
Refs.~\cite{khodas07,imambekov08}. These exponents differ from those predicted by  
Luttinger liquid theory raising the question whether the different results are 
compatible with each other. We address this question in Sec.~\ref{sec:apprexpr} of this 
paper, where we show that the results can be reconciled by taking appropriate limits. 
The apparent difference between the edge exponents valid along the dispersion curves and 
the Luttinger liquid result in the limit of vanishing energy can be traced back to the 
fact that the dispersion relations are curved and not straight, as is presumed by 
Luttinger liquid theory.

The exact values of the exponents found in Refs.~\cite{khodas07,imambekov08} are of 
importance; however, they are not sufficient for practical estimations of the DSF as 
long as the prefactors are not known. In this paper we construct an approximate formula 
for the DSF \cite{noteconf} based on the exponents of Refs.~\cite{khodas07,imambekov08}. 
Within the proposed scheme, the prefactor can be found using the well-known $f$-sum rule 
(see, e.g.~\cite{pitaevskii03:book}.) The result turns out to be consistent with 
numerical results by Caux and Calabrese \cite{caux06}. Besides, it is compatible with 
the results of Luttinger liquid theory \cite{haldane81,castro_neto94,astrakharchik04} and perturbation theory  \cite{brand05}. 
The approximate formula, in effect, takes into account single quasiparticle-quasihole 
excitations but neglects multiparticle excitations.  We also present an approximate 
expression for the static structure factor  and for the density-density correlation 
function, which is derived from the approximation for the DSF.

\section{Exact results for dynamic structure factor in Lieb-Liniger model}
\label{sec:exact}

We model cold bosonic atoms in a waveguide-like micro trap by a simple 1D gas of $N$
bosons with point interactions of strength $g_{\rm B}>0 $
\begin{equation}
H =  \sum_{i=1}^N -\frac{\hbar^2}{2 m}\frac{\partial^2}{\partial x_i^2}
+ g_{\text{B}} \sum_{1\leqslant i<j\leqslant N} \delta(x_i - x_j) \label{LLham}
\end{equation}
and impose periodic boundary conditions on the wave functions. The strength of 
interactions can be measured in terms of dimensionless parameter $\gamma= m 
g_{\text{B}}/(\hbar^2 n)$. In the limit of large $\gamma$, the model is known as the 
Tonks-Girardeau (TG) gas. In this limit, it can be mapped onto an ideal \emph{Fermi} gas 
since infinite contact repulsions emulate the Pauli principle. In the opposite limit of 
small $\gamma$, we recover the Bogoliubov model of weakly interacting bosons.

\subsection{DSF expansion in $1/\gamma$}

For finite $\gamma$, the model can also be mapped onto a Fermi gas \cite{cheon99} with 
local interactions, inversely proportional to $g_{\text{B}}$ 
\cite{girardeau04,granger04,brand05,cherny06}. Using the explicit form of the 
interactions, one can develop the time-dependent Hartree-Fock scheme 
\cite{brand05,cherny06} in the strong-coupling regime with small parameter $1/\gamma$. 
The scheme yields the correct expansion of the DSF up to the first order 
\cite{brand05,cherny06}
\begin{equation}
S(k,\omega)\frac{\ef}{N}= \frac{\kf}{4 k}\left(1+\frac{8}{\gamma}\right)
+\frac{1}{2\gamma}\ln \frac{\omega^{2}-\omega_{-}^{2}} {\omega_{+}^{2}-\omega^{2}}+
O\left(\frac{1}{\gamma^2}\right), \label{DSFlinear}
\end{equation}
for $\omega_{-}(k)\leqslant\omega\leqslant\omega_{+}(k)$, and zero elsewhere 
\cite{note1}. The symbol $O(x)$ denotes terms of order $x$ or even smaller. Here 
$\omega_\pm(k)$ are  the limiting dispersions \cite{note} that bound 
quasiparticle-quasihole excitations (see Fig.~\ref{fig:omplmi}); in the strong-coupling 
regime they take the form
\begin{equation}
\omega_\pm(k)={\hbar |2 \pf k \pm k^2|}(1-4/\gamma)/{(2 m)} +O(1/\gamma^2).
\label{ompmstrong}
\end{equation}
By definition, $\kf\equiv\pi n$ and $\ef\equiv\hbar^{2}\kf^{2}/(2m)$ are the Fermi wave 
vector and energy of TG gas, respectively.

\subsection{Link to Luttinger liquid theory}
\label{LLth}

Luttinger liquid theory describes the behavior of the DSF at low energies for arbitrary 
strength of interactions \cite{haldane81,astrakharchik04}. In particular, one can show 
\cite{astrakharchik04,castro_neto94} that in the vicinity of ``umklapp" point ($k=2\pi 
n$, $\omega =0$) it is given by
\begin{equation}
\frac{S(k,\omega)}{N}=\frac{n c}{\hbar\omega^{2}}
\left(\frac{\hbar\omega}{m c^{2}}\right)^{2K}
A(K)\left(1-\frac{\omega^{2}_{-}(k)}{\omega^2}\right)^{K-1}
\label{pitdsf}
\end{equation}
for $\omega\geqslant\omega_{-}(k)$, and zero otherwise. Within the Luttinger-liquid 
theory, the dispersion is \emph{linear} near the umklapp point: $\omega_{-}(k)\simeq 
c|k-2 \pi n|$. By definition, 
\begin{equation}
K\equiv \hbar\pi n/(m c)
\label{Kdef}
\end{equation}
and $c$ is the sound velocity. For the repulsive bosons, the value of parameter $K$ lies 
between $1$ (TG gas) and $+\infty$ (ideal Bose gas). In the strong-coupling regime, the 
linear behavior of the dispersions (\ref{ompmstrong}) at small momentum determines the 
sound velocity, which allows us to calculate the value of the Luttinger parameter
\begin{equation}
K=1 +4/\gamma + O(1/\gamma^2).
\label{Kstrong}
\end{equation}
The coefficient $A(K)$ is model-dependent; in the Lieb-Liniger model, it is known in two 
limiting cases: $A(K)=\pi/4$ at $K=1$ and $A(K)\simeq 
8^{1-2K}\exp(-2\gamma_{\mathrm{c}}K)\pi^{2}/\Gamma^{2}(K)$ for $K\gg 1$ 
\cite{astrakharchik04}, where $\gamma_{\mathrm{c}}=0.5772\ldots$ is the Euler constant 
and $\Gamma(K)$ is the gamma function.

By comparing the first-order expansion (\ref{DSFlinear}) in the vicinity of the umklapp 
point with Eq.~(\ref{pitdsf}) and using the expansion (\ref{Kstrong}), one can easily 
obtain the model-dependent coefficient at large but \emph{finite} interactions when  
$K-1\ll 1$
\begin{align}
\label{akseries}
A(K)
=\frac{\pi}{4}[1 - \left(1+4\ln 2\right)(K-1)]
 + O\left((K-1)^2\right).
\end{align}

Note that the relation (\ref{pitdsf}) leads to different exponents precisely at the 
umklapp point and outside of it:
\begin{equation}
\label{pitdsfexp} S(k,\omega)\sim \left\{\begin{array}{ll}
\omega^{2(K-1)},& k=2\pi n,\\
(\omega-\omega_{-})^{K-1},& k\not=2\pi n.
\end{array}\right.
\end{equation}

\subsection{Exact edge exponents from the Lieb-Liniger solutions}

\begin{figure}[t,b]
\includegraphics[width=.8\columnwidth]{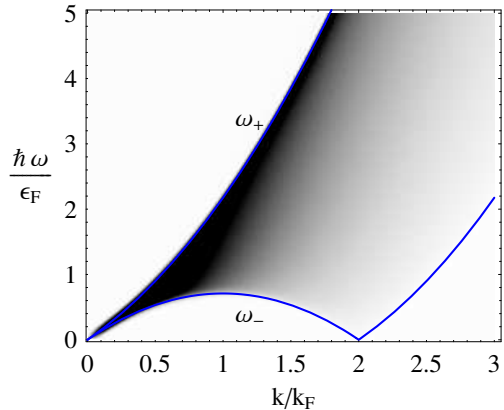}
\caption{\label{fig:omplmi} (Color online) Numerical values of the DSF 
(\ref{eqn:dsfenergy}) for the coupling parameter $\gamma = 10$ \cite{caux06}. The 
dimensionless value of the rescaled DSF $S(k,\omega)\varepsilon_{F}/N$ is shown in 
shades of gray between zero (white) and 1.0 (black). The upper and lower solid (blue) 
lines represent the dispersions $\omega_+(k)$ and $\omega_-(k)$, respectively, limiting 
the single ``particle-hole" excitations in the Lieb-Liniger model at $T=0$. The 
dispersions are obtained numerically by solving Lieb-Liniger's system of integral 
equations (see Appendix \ref{sec:LL}). The gray scale plot of the DSF demonstrates that 
the main contribution to the DSF comes from the single particle-hole excitations, lying 
inside the region $\omega_-(k)\leqslant\omega\leqslant\omega_+(k)$ (see also 
Fig.~\ref{fig:dsf}). }
\end{figure}

\begin{figure}[t,b]
\includegraphics[width=.6\columnwidth]{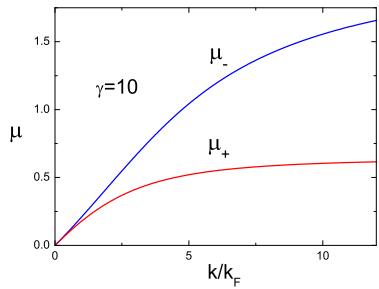}
\caption{\label{muplmi} (Color online) Typical behavior of the exact exponents in 
Eq.~(\ref{glazexp}). The diagram shows $\mu_\pm$ for $\gamma=10$ obtained numerically 
using the method of Ref.~\cite{imambekov08} described in Appendix \ref{sec:LL}. }
\end{figure}

As was shown in Refs.~\cite{khodas07,imambekov08} (see also \cite{cheianov08}), within 
the Lieb-Liniger model the DSF exhibits the following power-law behavior near the 
borders of the spectrum $\omega_\pm(k)$
\begin{equation}
S(k,\omega)\sim \big|\omega-\omega_{\pm}(k)\big|^{\mp \mu_{\pm}(k)}. \label{glazexp}
\end{equation}
The positive exponents $\mu_\pm$ \cite{note} are related to the quasi-particle 
scattering phase and can be calculated in the thermodynamic limit by solving a system of 
integral equations \cite{imambekov08}. In particular, Imambekov and Glazman 
\cite{imambekov08} found the following right limit
\begin{equation}
\label{mumiumklapp} 
\lim_{k\to 2 \pi n^-}\mu_{-}(k)=2\sqrt{K}(\sqrt{K}-1),
\end{equation}
which is different from the Luttinger liquid exponent (\ref{pitdsfexp}). However, 
Imambekov's and Glazman's result (\ref{mumiumklapp}) is accurate in the immediate 
vicinity of $\omega_{-}$ provided that the finite curvature of $\omega_{-}(k)$ is taken 
into consideration. Thus the difference in the exponents can be attributed 
\cite{imambekov08} to the linear spectrum approximation within the Luttinger liquid 
theory. Note, however, that the thin ``strip" in $\omega$-$k$ plane, where the exponents 
are different, vanishes in the point $k=2\pi n$; hence, the Luttinger exponent $2(K-1)$ 
becomes exact there.

A typical behavior of the exponents is shown in Fig.~\ref{muplmi}. As described in 
Appendix \ref{sec:LL}, the exponents can be easily evaluated by solving equation 
(\ref{LLshitTL}) for the shift function and using Eq.~(\ref{IGmu}).

\subsection{Algebraic Bethe ansatz}

\begin{figure}[t,b]
\begin{center}
\noindent\includegraphics[width=\columnwidth]{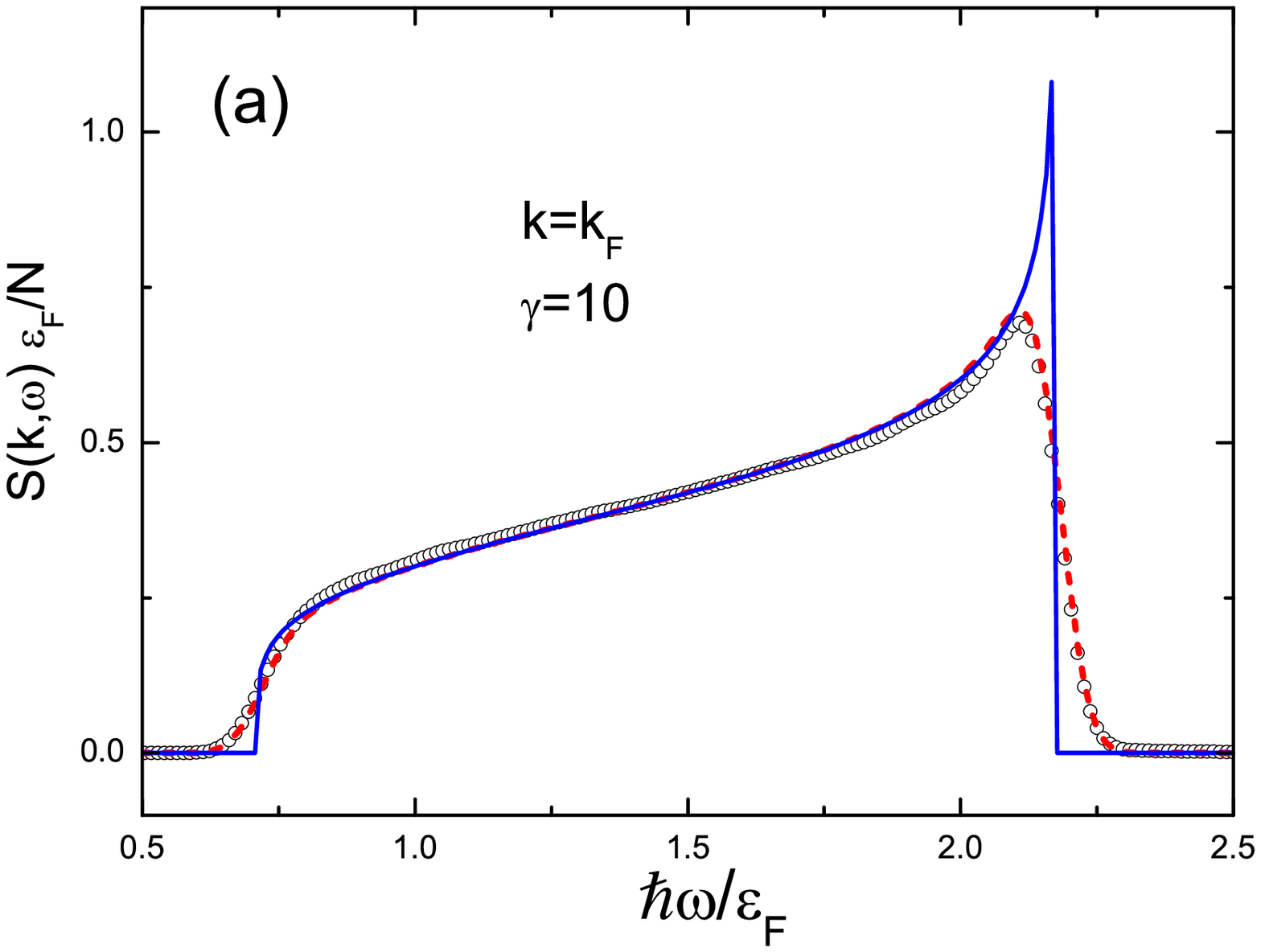}\\
\includegraphics[width=\columnwidth]{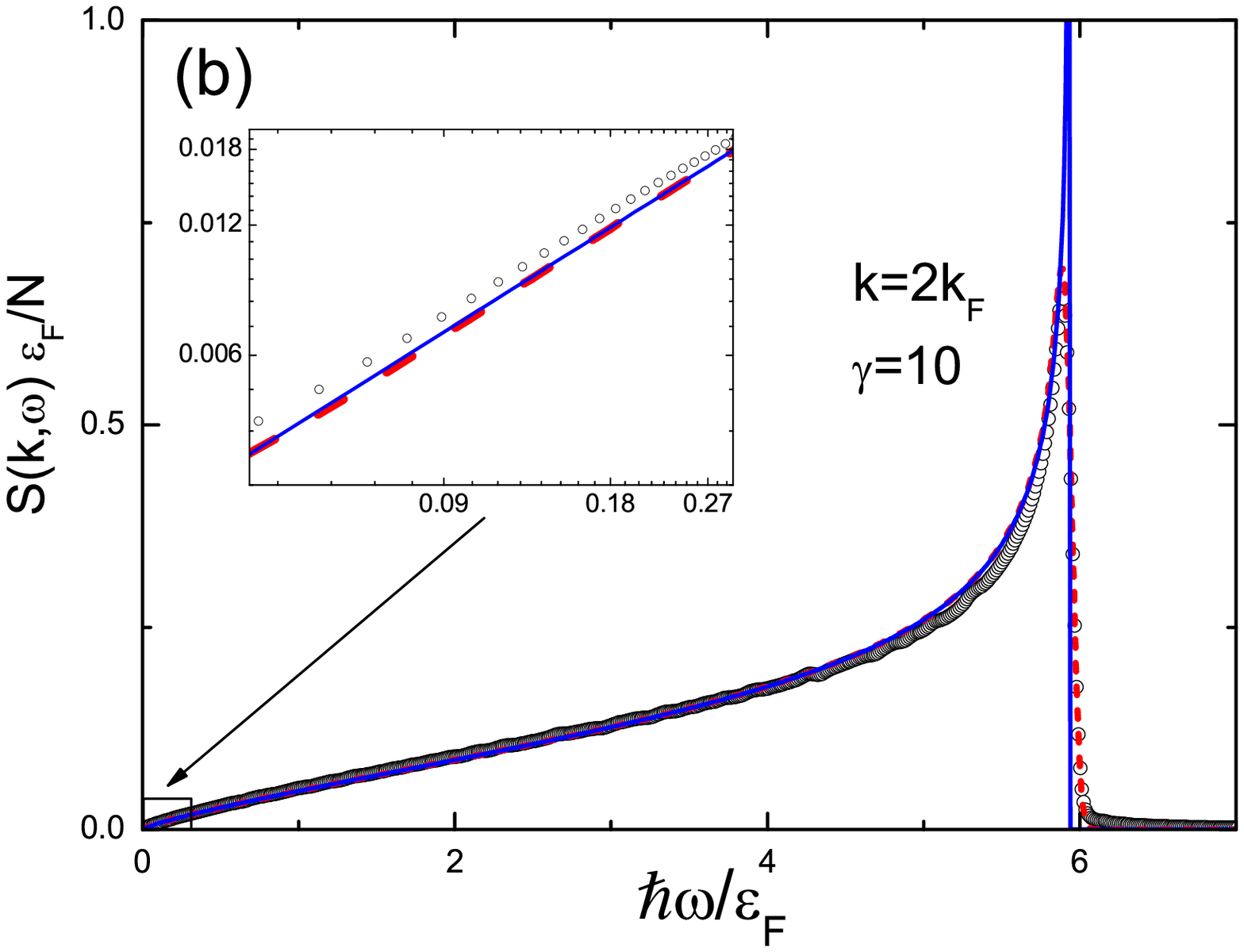}
\end{center}
\caption{\label{fig:dsf} (Color online) The Dynamic Structure Factor (DSF) in the 
thermodynamic limit. The proposed approximation (\ref{dsfapp1}) (line) is compared to 
numerical data from Caux and Calabrese \cite{caux06} (open dots). The dashed (red) line 
shows the data of Eq.~(\ref{dsfapp1}) convoluted in frequency with a Gaussian of width 
$0.07\ef/\hbar$ in order to simulate smearing that was used in generating the numerical 
results of Ref.~\cite{caux06}. The numerical data of Ref.~\cite{caux06} suggest that 
contributions from multi-particle excitations for $\omega>\omega_+$ (sharp line in parts 
a and b) are very small. Such contributions are not accounted for by the formula 
(\ref{dsfapp1}). Insert: DSF at the umklapp point in logarithmic scale. The graph shows 
that the DSF behaves as predicted by the Luttinger liquid theory (\ref{glazdsfexp}) with 
the exponent $2(K-1)$, where $K=1.402\ldots$ at $\gamma=10$.}
\end{figure}

Recent progress in the computation of correlation functions within the Lieb-Liniger 
model and other 1D models has been achieved through the algebraic Bethe ansatz 
\cite{caux06}. In this method, matrix elements of the density operator involved in 
Eq.~(\ref{eqn:dsfenergy}) were calculated with the algebraic Bethe ansatz. They are 
given by the determinant of a matrix, which can be evaluated numerically for a finite 
number of particles. So, this method is based on combining integrability and numerics. 
The results of the numerical calculations of Ref.~\cite{caux06} are shown in 
Figs.~\ref{fig:omplmi} and \ref{fig:dsf}.

\section{Approximate expression for dynamic structure factor}

\subsection{Approximate expression for arbitrary values of interaction strength}
\label{sec:apprexpr}

Here we suggest a phenomenological expression, which is consistent with all the 
above-mentioned results. It reads
\begin{equation}
S(k,\omega)=C \frac{(\omega^{\alpha}-\omega_{-}^{\alpha})^{\mu_{-}}}
{(\omega_{+}^{\alpha}-\omega^{\alpha})^{\mu_{+}}}
\label{dsfapp1}
\end{equation}
for $\omega_{-}(k)\leqslant\omega\leqslant\omega_{+}(k)$, and zero otherwise. It follows 
from energy and momentum conservation that $S(k,\omega)$ is exactly equal to zero below 
$\omega_{-}(k)$ for $0\leqslant k \leqslant 2 \pi n$. In the other regions of $\omega > 
\omega_{+}$ and  $\omega < \omega_{-}$ (for $k > 2 \pi n$), possible contributions can 
arise due to coupling to multi-particle excitations \cite{lieb63:2}. However, these 
contributions are known to vanish in the Tonks-Girardeau ($\gamma \to \infty$) and 
Bogoliubov ($\gamma \to 0$) limits and are found to be very small numerically for finite 
interactions \cite{caux06}.

In Eq.~(\ref{dsfapp1}) $C$ is a normalization constant, $\mu_{+}(k)$ and $\mu_{-}(k)$ 
are the exponents of Eq.~(\ref{glazexp}), and $\alpha\equiv 1+1/\sqrt{K}$. From the 
definition of $K$ (\ref{Kdef}), one can see that for repulsive spinless bosons 
$K\geqslant 1$, and, hence, $1<\alpha\leqslant 2$. The normalization constant depends on 
the momentum but not the frequency and can be determined from the $f$-sum rule 
\cite{pitaevskii03:book}
\begin{equation}
\int_{0}^{+\infty} \d\omega\, \omega S(k,\omega)= N\frac{k^{2}}{2m}. 
\label{fsum}
\end{equation}
In Eq.~(\ref{dsfapp1}) we assume that the value of the exponent $\mu_{-}(k=2\pi n)$ 
coincides with its limiting value (\ref{mumiumklapp}) in vicinity of the umklapp point.

The most general way of obtaining $\omega_{\pm}(k)$, $\mu_{\pm}(k)$, and $K$ is to solve 
numerically the corresponding integral equations (see Appendix \ref{sec:LL}).

Note that the sum rule for the isothermal compressibility \cite{pitaevskii03:book}
\begin{equation}
\label{m-1} \lim_{k\to 0}\int_{0}^{+\infty} \frac{S(k,\omega)\,\d\omega}{N
\omega}=\frac{1}{2 mc^2}
\end{equation}
is satisfied by virtue of Eq.~(\ref{relexpzero}) and the phonon behavior of the 
dispersions at small momentum: $\omega_{\pm}(k)\simeq c k$ (see Fig.~\ref{fig:omplmi}).

Now one can see from (\ref{dsfapp1}) that
\begin{equation}
\label{glazdsfexp} S(k,\omega)\sim \left\{\begin{array}{ll}
\omega^{2(K-1)},& k=2\pi n,\\
(\omega-\omega_{-})^{\mu_{-}(k)},& k\not=2\pi n.
\end{array}\right.
\end{equation}
Thus, the suggested formula (\ref{dsfapp1}) is consistent with both the Luttinger liquid 
behavior at the umklapp point and Imambekov's and Glazman's power-law behavior in 
vicinity of it, as it should be.

In the strong-coupling regime, Eq.~(\ref{dsfapp1}) yields the correct first order 
expansion (\ref{DSFlinear}). In order to show this, it is sufficient to use the 
strong-coupling values of $K$ (\ref{Kstrong}), the exponents (\ref{mustrong}), and the 
frequency dispersions (\ref{ompmstrong}).

Comparison with the numerical data by Caux and Calabrese \cite{caux06} 
(Fig.~\ref{fig:dsf}) shows that the suggested formula works well in the regimes of both 
weak and strong coupling. 

Let us discuss how the Bogoliubov approximation arises in the weak-coupling regime in 
spite of the absence of the Bose-Einstein condensation in one dimension even at zero 
temperature \cite{book:bogoliubov70,hohenberg67}. At small $\gamma$, the upper 
dispersion curve $\omega_{+}(k)$ is described well \cite{lieb63:2} by the Bogoliubov 
relation \cite{bogoliubov47}
\begin{equation}
\hbar\omega_k=\sqrt{T_k^{2}+4 T_k \ef \gamma/\pi^{2}},
\label{bogdisp}
\end{equation}
where $T_{k}=\hbar^{2}k^{2}/(2m)$ denotes the usual one-particle kinetic energy. 
Besides, when $q$ is finite and $\gamma\to0$, the associated exponents $\mu_{+}$ 
approach the limiting value (\ref{relexp}), which in turn is very close to one. This 
implies that the DSF has a strong singularity near $\omega_{+}$, and, hence, it is 
localized almost completely within a small vicinity of the upper branch (see 
Fig.~\ref{fig:approx2}). Thus, the behavior of the DSF simulates the $\delta$-function 
spike. One can simply put $S_{\mathrm{Bog}}(k,\omega)=C \delta(\omega-\omega_{k})$ and 
determine the constant $C$ from the $f$-sum rule (\ref{fsum})
\begin{equation}
S_{\mathrm{Bog}}(k,\omega)=N\frac{T_{k}}{\hbar\omega_k} \delta(\omega-\omega_{k}).
\label{dsfbog}
\end{equation}

\subsection{Simplified analytic approximation for intermediate
and large strength of interactions}

\begin{figure}[t,b]
\begin{center}
\noindent\includegraphics[width=\columnwidth]{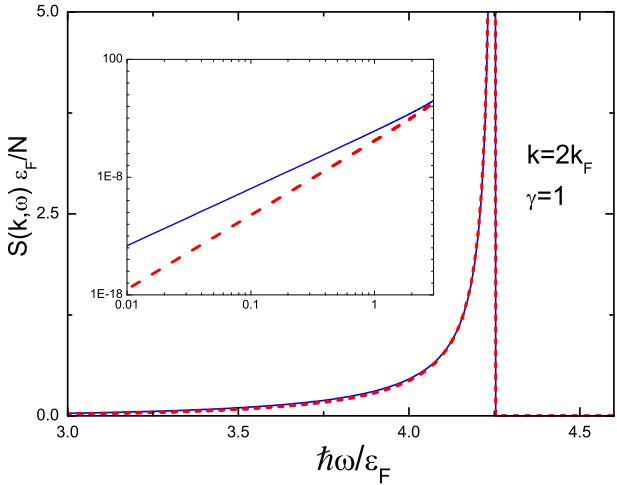}\\
\end{center}
\caption{\label{fig:approx2} (Color online) Comparison of the two approximations for the 
DSF. The solid (blue) line represents the ``universal" approximation (\ref{dsfapp1}). The 
dashed (red) line is the approximation  (\ref{dsfc}). The two curves coincide almost 
everywhere except for the umklapp point ($\omega=0$, $k=2\pi n$). The ``universal" 
approximation reproduces the correct power-low behavior of the Luttinger liquid theory 
$S(k,\omega)\sim \omega^{2(K-1)}$ with $K=3.425\ldots$ at $\gamma=1$, see the insert. 
However, the difference in absolute values is negligible due to a strong suppression of 
the DSF outside the close vicinity of the upper brunch.}
\end{figure}

One can further simplify the expression for the DSF and replace the parameter $\alpha$ 
in Eq.~(\ref{dsfapp1}) by its limiting value $\alpha=2$ for the Tonks-Girardeau gas, 
which turns out to be a good approximation even for intermediate coupling strength 
$\gamma\gtrsim 1$. This replacement allows us to write down  the normalization constant 
explicitly. From the $f$-sum rule we obtain
\begin{align}
S(k,\omega)= &N\frac{k^{2}}{m}\frac{\Gamma(2+\mu_{+}-\mu_{-})}
{\Gamma(1+\mu_{-})\Gamma(1-\mu_{+})} \frac{(\omega^{2}-\omega_{-}^{2})^{\mu_{-}}}
{(\omega_{+}^{2}-\omega^{2})^{\mu_{+}}}\nonumber \\
&\times (\omega_{+}^{2}-\omega_{-}^{2})^{\mu_{+}-\mu_{-}-1}
\label{dsfc}
\end{align}
for $\omega_{-}(k)\leqslant\omega\leqslant\omega_{+}(k)$, and zero otherwise. This 
approximation ensures all the properties of the DSF mentioned in Sec.~\ref{sec:exact}, 
except for the Luttinger liquid theory predictions in the close vicinity of the umklapp 
point (see discussion in Sec.~\ref{LLth}). However, outside the umklapp point, it agrees 
well with the Caux and Calabrese numerical data (see Figs.~\ref{fig:dsf} and 
\ref{fig:approx2}).

\begin{figure}[t,b]
\noindent\includegraphics[width=.9\columnwidth]{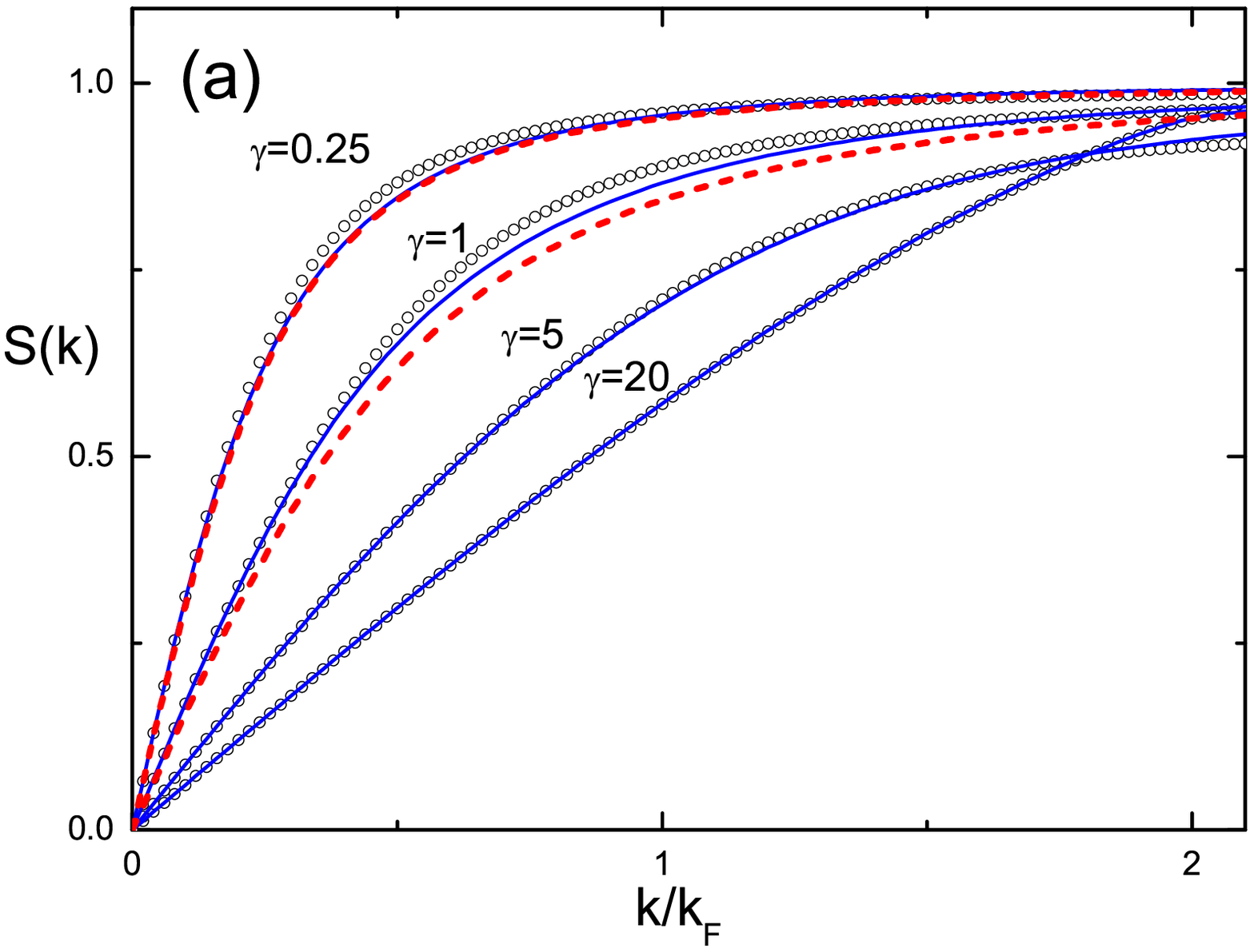}\\
\includegraphics[width=.9\columnwidth]{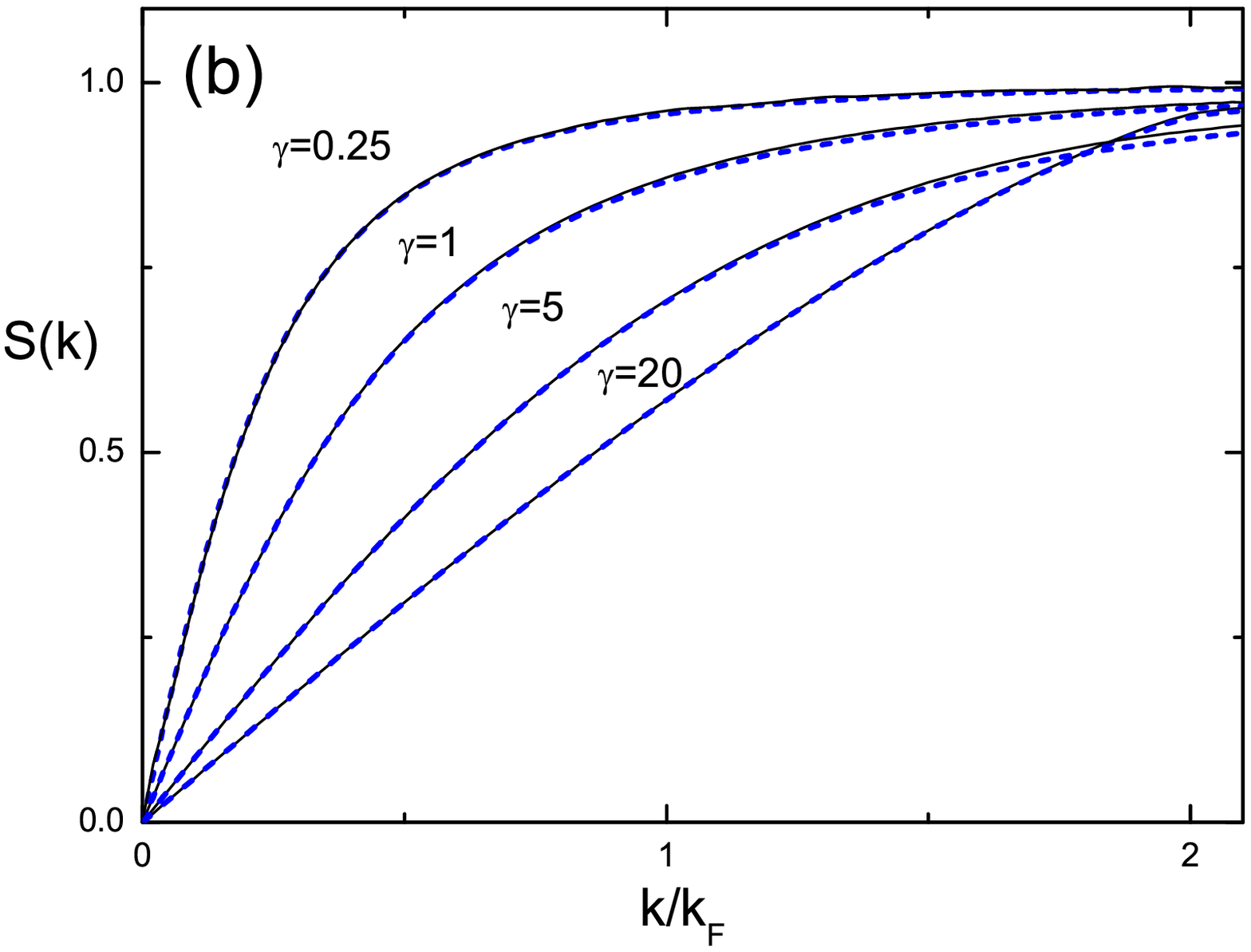}
\caption{\label{fig:ssf} (Color online) The static structure factor versus wavenumber 
for different values of the coupling constant $\gamma$. (a) The numerical data by Caux 
and Calabrese \cite{caux06} (open circles) are compared with the proposed analytical 
formula (\ref{ssf}) (solid lines). The dashed (red) line shows the static structure 
factor in the Bogoliubov limit (\ref{ssfbog}). (b) The static structure factor obtained 
with Eq.~(\ref{ssfgen}) from the general formula for the DSF (\ref{dsfapp1}) is shown by 
the solid line. These data are consistent with the analytical formula (\ref{ssf}) 
(dashed line). This indicates that the analytical formula  for the static structure 
factor can be used even for small values of $\gamma$. }
\end{figure}

From the explicit formula (\ref{dsfc}) one can find analytic expressions for the static 
structure factor and the dynamic polarizability. The static structure factor $S(k)\equiv 
\langle\hat{\rho}_{k} \hat{\rho}_{-k} \rangle/N$ contains information about the static 
correlations of the system and it is directly related to the pair distribution function 
\cite{pitaevskii03:book,pines66:book}
\begin{equation}
g(x)=1+\int_{0}^{+\infty} \frac{\d k}{\pi n}\,\cos(k x)\big[S(k)-1\big].
\label{sgx}
\end{equation}
The static structure factor can be obtained by integrating the DSF over the frequency
\begin{align}
S(k)= \frac{\hslash}{N}\int_{0}^{+\infty}S(k,\omega)\,\d\omega.
\label{ssfgen}
\end{align}
Note that the ``phonon" behavior of both dispersions ensures the correct behavior of the 
static structure factor at small momentum. Indeed, it follows from the general 
expression (\ref{dsfapp1}) that $S(k)\simeq\hbar k/(2mc)$. In the large-momentum limit, 
we have $\omega_{+}/\omega_{-}\simeq 1$, which leads to the correct asymptotics $S(k)\to 
1$ as $k\to+\infty$. Equations (\ref{dsfc}) and (\ref{ssfgen}) yield
\begin{align}
S(k)=&{}_{2}F_{1}\Big(\frac{3}{2}\!+\!\mu_{-}\!-\!\mu_{+},1\!+\!\mu_{-},2\!+\!\mu_{-}\!-\!\mu_{+},1\!-\!\frac{\omega_{-}^{2}}{\omega_{+}^{2}}\Big) \nonumber \\
&\times\frac{\hbar k^{2}}{2m\omega_{+}}\Big(\frac{\omega_{-}}{\omega_{+}}\Big)^{1+2\mu_{-}},
\label{ssf}
\end{align}
where ${}_{2}F_{1}$ is the hypergeometric function. The results for the static structure 
factor are plotted in Fig.~\ref{fig:ssf}. One can see that the formula for the static 
structure function works well even for weak coupling. This is due to the smallness of 
the DSF contribution to the static structure function at the umklapp point for small 
$\gamma$. Thus, the approximate formula provides a good accuracy for arbitrary strength 
of interactions. In the weak-coupling regime, one can obtain a good approximation for 
the static structure function from the Bogoliubov formula (\ref{dsfbog}) for the DSF
\begin{equation}
S(k)=\frac{T_{k}}{\hbar\omega_k}.
\label{ssfbog}
\end{equation}
 
\begin{figure}[t,b]
\noindent\includegraphics[width=.9\columnwidth]{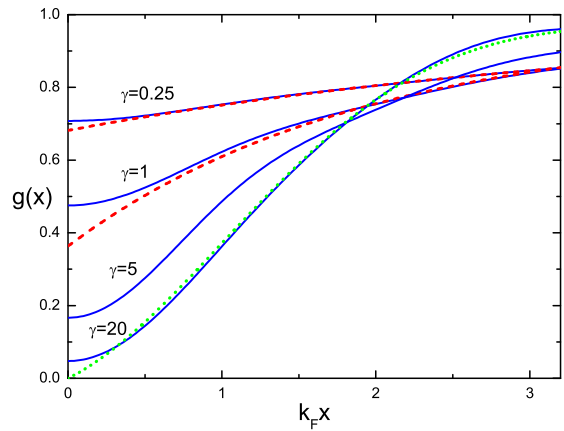} 
\caption{\label{fig:gx} (Color online) The pair distribution function $g(x)$ versus the 
distance in dimensionless units $1/\kf$ ($\kf\equiv\pi n$) for different values of the 
coupling constant $\gamma$. The solid (blue) line represents the pair distribution 
function (\ref{sgx}) obtained using the approximation (\ref{ssf}). The Bogoliubov 
approximation (\ref{ssfbog}) is indicated by dashed (red) line, while the 
strong-coupling approximation (\ref{gxfirstour}) by  dotted (green) line. The values of 
$g(0)$ are consistent with the results of Ref.~\cite{gangardt03}.}
\end{figure}

The behavior of the pair distribution function (\ref{sgx}) in the Lieb-Liniger model was 
studied at large \cite{korepin93:book} and short distances 
\cite{gangardt03,cherny06,sykes08} in various regimes. For $\gamma\ll 1$, one can obtain 
from Eqs.~(\ref{sgx}) and (\ref{ssfbog}) the analytical expression \cite{sykes08}
\begin{equation}
g(x)=1-\sqrt{\gamma}\big[\bm{L}_{-1}(2\sqrt{\gamma}\,\kf x/\pi)
-I_1(2\sqrt{\gamma}\,\kf x/\pi)\big],
\label{gxbog}
\end{equation}
where $\bm{L}_{-1}(x)$ is the modified Struve function and $I_1(x)$ is a Bessel 
function. In the opposite limit $\gamma\gg 1$, one can directly use the strong-coupling 
expression (\ref{DSFlinear}) for the DSF and obtain \cite{cherny06}
\begin{align}
g&(x)=\ 1-\frac{\sin^{2}z}{z^{2}}
-\frac{2\pi}{\gamma}\frac{\partial}{\partial z}\frac{\sin^{2}z}{z^{2}}
-\frac{4}{\gamma}\frac{\sin^{2}z}{z^{2}}\nonumber\\
& +\frac{2}{\gamma}\frac{\partial}{\partial z}
\left[\frac{\sin z}{z}\int_{-1}^{1}\d\eta\,\sin(\eta z)
\ln\frac{1+\eta}{1-\eta}\right] + O(\gamma^{-2}),
\label{gxfirstour}
\end{align}
where $z=\kf x=\pi n x$. 
The last equation implies that $g(x=0)$ vanishes not only in the TG limit but also in 
the first order of $\gamma^{-1}$, which is consistent with the results of 
Refs.~\cite{lieb63:1,gangardt03}.

The behavior of $g(x)$, obtained from the formula (\ref{ssf}) is shown in 
Fig.~\ref{fig:gx}. It is consistent with both the weak- and the strong-coupling limits. 

The dynamic polarizability determines the linear response of the density  to an external 
field \cite{pitaevskii03:book,pines66:book}. It can be calculated using the DSF
\begin{align}
\chi(k,z)=\int_{0}^{+\infty}\frac{2\omega'S(k,\omega')}{{\omega'}^{2}-z^{2}}\d\omega'
\label{chigen}
\end{align}
On substituting Eq.~(\ref{dsfc}) in (\ref{chigen}), we get
\begin{align}
\chi(k,z)=&{}_{2}F_{1}\Big(1,1+\mu_{-},2+\mu_{-}\!-\mu_{+},\frac{\omega_{+}^{2}-\omega_{-}^{2}}{z^{2}-\omega_{-}^{2}}\Big)\nonumber \\
&\times N\frac{k^{2}}{m}\frac{1}{\omega_{-}^{2}-z^{2}}.
\label{chi}
\end{align}
For a retarded response, we should put here $z=\omega+i\varepsilon$. At zero temperature 
the relation $S(k,\omega) =\im \chi(k,\omega +i\varepsilon)/\pi$ holds.

The obtained relations (\ref{ssf}) and (\ref{chi}) successfully reproduce the 
Tonks-Girardeau limit, considered in detail in Refs.~\cite{brand05,cherny06}.

\section{Conclusion}

We have discussed an approximate formula [Eq.\ (\ref{dsfapp1})] for the DSF of the 
one-dimensional Bose gas at zero temperature, which can be used for a wide range of 
momenta, energies, and coupling strengths. It neglects, in effect, only the 
multiparticle excitations, whose contribution is small, anyway, outside the bounds given 
by the dispersion curves $\omega_\pm$. Our formula is consistent with the predictions of 
the Luttinger liquid theory. It gives the exact exponents at the edge of the spectrum, 
the correct first-order expansion in the strong-coupling regime, and shows good 
agreement with the available numerical data. For intermediate and large values of the 
interaction strength $\gamma\gtrsim 1$ and outside the close vicinity of the umklapp 
point ($\omega=0$, $k=2\pi n$), the further simplified analytic formulas for the DSF 
(\ref{dsfc}) and the dynamic polarizability (\ref{chi}) provide excellent accuracy. The 
analytic expression (\ref{ssf}) for the static structure factor works well even for weak 
interactions.

Our results provide a reference against which experimental measurements of static and 
dynamic density correlations in the one-dimensional Bose gas can be tested. They further 
provide a basis for future work on the consequences of correlations in this interesting 
system.

\begin{acknowledgments}
The authors are grateful to Lev Pitaevskii for valuable discussions and to 
Jean-Sebastien Caux for making the data of numerical calculations of Ref.~\cite{caux06} 
available to us and to Thomas Ernst for checking our numerical results. JB is supported 
by the Marsden Fund Council (contract MAU0706) from Government funding, administered by 
the Royal Society of New Zealand. AYuCh thanks Massey University for hospitality.
\end{acknowledgments}

\appendix

\section{The Lieb-Liniger solutions and the edge exponents}
\label{sec:LL}

For the Lieb-Liniger model (\ref{LLham}) of $N$ bosons embedded in the circuit of length 
$L$, the Bethe ansatz yields \cite{lieb63:1,yang69,korepin93:book} the following system 
of equations for $N$ quasimomenta $\lambda_j$
\begin{equation}
L \lambda_j + \sum_{k=1}^N \theta (\lambda_j-\lambda_k)=2\pi n_j
\label{LLnu}
\end{equation}
where $\theta(\lambda)\equiv 2 \arctan[\hbar^2 \lambda/(g_{\mathrm B} m)]$, and $n_j$ 
are integer for odd $N$ and half-integer for even $N$. In the same way as for the gas of 
non-interacting fermions, eigenstates of Hamiltonian (\ref{LLham}) are completely 
described by a certain set of $N$ numbers $n_j$; the total energy and momentum of the 
states are given by
\begin{equation}
E_{N} =\frac{\hbar^{2}} {2m} \sum_{j=1}^{N} \lambda_j^{2}, \quad P_{N} =\hbar
\sum_{j=1}^{N} \lambda_j=\frac{2\pi\hbar}{L} \sum_{j=1}^{N} n_j.
\label{enmomLL}
\end{equation}
One can also write down explicitly \cite{lieb63:1,yang69,korepin93:book} rather 
cumbersome expression for the eigenfunctions in terms of quasimomenta. One can prove the 
following properties of the Lieb-Liniger solutions
\cite{gaudin83book,korepin93:book}: \\
\emph{i)} The Bethe ansatz solutions form a complete set of eigenfunctions.\\
\emph{ii)} There is one-to-one correspondence between $n_j$ and $\lambda_j$; if $n_j>n_k$
then $\lambda_j>\lambda_k$. \\
\emph{iii)} The set of numbers $n_j=j-1 -(N-1)/2$, $1 \leqslant j\leqslant N$ provides the 
ground state; for the associated quasimomenta we have $|\lambda_j|\leqslant q_0\leqslant 
\pi n$.\\
\emph{iv)} In the thermodynamic limit, when $N\to\infty$ and $n=N/L=\mathrm{const}$, the 
distance between the 
consecutive quasimomenta tends to zero: $\lambda_j-\lambda_k\sim 1/L$.

In order to pass to thermodynamic limit, it is convenient to introduce the density of 
quasimomenta $\rho(\lambda)\equiv\lim 1/[L(\lambda_{i+1}-\lambda_{i})]$. From 
Eq.~(\ref{LLnu}) and property \emph{iii)} one can obtain \cite{lieb63:1} a system of 
linear integral equations for the density in the ground state
\begin{align}
  2\pi \rho(\nu)&= 1+\int_{-q_0}^{q_0} K(\nu,\mu)\rho(\mu)\,d\mu, \label{LLGSrho}\\
  n &=\int_{-q_0}^{q_0} \rho(\nu)\,d\nu.
\label{LLGSn}
\end{align}
We denote 
\begin{equation}
K(\nu,\mu)\equiv\frac{2g_{\mathrm{B}}m}{\hbar^2}
\left[\left(\frac{g_{\mathrm{B}}m}{\hbar^2}\right)^2+(\nu-\mu)^2\right]^{-1}
\label{Kker}
\end{equation}
and assume that quasimomenta fill ``Fermi segment" with bounds $\pm q_0$. The density at 
the bounds can be related \cite{korepin93:book} to the Luttinger liquid parameter by 
equation $\rho(\pm q_{0})=\sqrt{K}/(2\pi)$.

Classification of all the excitations can be done in the same manner as for the 
Tonks-Girardeau gas of non-interacting fermions ($g_{\mathrm B}\to+\infty$). Thus, in 
order to create elementary particle-like excitation, one needs to add a quasimomentum 
$\lambda_p$ beyond ``Fermi segment" $[-q_0,q_0]$. By contrast, for a hole-like 
excitation, one needs to remove a quasimomentum lying inside the ``Fermi segment". All 
the excitations can be constructed from the above elementary excitations. Adding or 
removing a particle with quasimomentum $\lambda_p$ leads to a new set of quasimomenta 
$\tilde{\nu}_{j}$. One can conveniently define a shift function 
$F_B(\nu_j|\lambda_p)\equiv\pm (\nu_j-\tilde \nu_j)/(\nu_{j+1}-\nu_{j})$, where we take 
plus for a particle-like excitation and minus for a hole-like excitation. It follows 
from (\ref{LLnu}) that the shift function obeys the integral equation in the 
thermodynamic limit \cite{korepin93:book,imambekov08}
\begin{equation}
F_B(\nu|\lambda)-\frac{1}{2\pi}\int_{-q_0}^{q_0}K(\nu,\mu)F_B(\mu|\lambda)\,d\mu=
\frac{\pi+\theta(\nu-\lambda)}{2\pi}.
\label{LLshitTL}
\end{equation}
Using Eq.l~(\ref{enmomLL}) and the definitions of the density and shift function, one 
can evaluate the frequency and the wavenumber of elementary single excitation in the 
thermodynamic limit \cite{korepin93:book}
\begin{align}
 \omega_{\mathrm{p,h}}(\nu)&= \pm \frac{\hbar}{2m}
\Big[\nu^2- \int_{-q_0}^{q_0}2\mu F_B(\mu|\nu)\,d\mu\Big], \label{energyelem} \\
  k_{\mathrm{p,h}}(\nu)&= \pm
\Big[\nu+ \int_{-q_0}^{q_0}\theta(\nu-\mu)\rho(\mu)\,d\mu\Big].
\label{momentumelem}
\end{align}
For the particle excitation, as discussed above, we should take $\nu>q_0$, and for the 
hole excitation $|\nu|<q_0$. Note that Eq.~(\ref{momentumelem}) yields $k_h(\pm 
q_0\mp0)=\mp\kf$ and $k_p(\pm q_0\pm0)=\pm\kf$, as it should be.

Once equations (\ref{LLGSrho}), (\ref{LLGSn}), and (\ref{LLshitTL}) are solved, we can 
write down the dispersions $\omega_{\pm}(k)$ \cite{note} bounding single particle-hole 
excitations (see Fig.~\ref{fig:omplmi}). The upper $\omega_{+}$ branch is constructed by 
creating hole with quasimomentum $q_0$ and particle with quasimomentum $\lambda>q_0$, 
while the lower $\omega_{-}$ branch by creating hole inside the ``Fermi 
segment"($|\lambda|<q_0$) and particle with quasimomentum lying just above $q_0$. 
Explicitly, the branches and associated momenta corresponds to the pair excitations 
$\lambda_h=\lambda_0$ and $\lambda_p=\lambda + \lambda_0$ (upper, or ``$+$", branch), 
$\lambda_h=\lambda_0-\lambda$ and $\lambda_p=\lambda_0$ (lower, or ``$-$",branch). 
Summing energies (\ref{energyelem}) and wavevectors (\ref{momentumelem}) of the 
corresponding elementary excitations yields the dependencies $\omega(\lambda)$ and 
$k(\lambda)$ for the upper branch
\begin{align}
\omega_+(\lambda)&= \omega_h(\lambda_0)+\omega_p(\lambda +\lambda _0),\nonumber\\
k_+(\lambda)&= k_h(\lambda_0)+ k_p(\lambda + \lambda_0),
\nonumber
\end{align}
and the lower branch
\begin{align}
\omega_-(\lambda)&= \omega_h(\lambda_0-\lambda)+\omega_p(\lambda_0),\nonumber\\
k_-(\lambda)&= k_h(\lambda_0-\lambda)+k_p(\lambda_0).
\nonumber
\end{align}
After excluding quasimomentum, we obtain the dispersions as functions of $k$.

It turns out \cite{korepin93:book} that the shift function in the thermodynamic limit is 
directly related to the scattering phase of two particles (holes) in the presence of 
filled ``Fermi segment". As was shown by Imambekov and Glazman \cite{imambekov08}, the 
edge exponents, specifying the behavior of the DSF near the dispersion curves, can be 
written down in terms of the associated phases $\delta_{\pm}(\lambda)=2\pi F_B(\pm 
q_0,\lambda)$
\begin{equation}
\mu_{\pm} =1 - \frac12\left(\frac{1}{\sqrt{K}}+\frac{\delta_{+}-\delta_-}
{2\pi}\right)^2-\frac12\left(\frac{\delta_{+}+\delta_-}{2\pi}\right)^2.
\label{IGmu}
\end{equation}
The exponents obey \cite{imambekov08} the limiting relations
\begin{align}
\mu_{\pm}(0)&=0,
\label{relexpzero}\\
\mu_{+}(+\infty)&=1-1/(2K).
\label{relexp}
\end{align}
In the strong-coupling limit, one can show \cite{khodas07} that
\begin{equation}
\mu_{\pm}(k)=2k/(\pi n\gamma)+ O(1/\gamma^2).
\label{mustrong}
\end{equation}


\begin{thebibliography}{39}
\bibitem[{not({\natexlab{b}})}]{note1}
\bibinfo{note}{It follows from the symmetry considerations that
  $S(k,\omega)=S(-k,\omega)$. For this reason, in this paper we assume that in
  the formulas for the dynamic and static stucture factors $k\geqslant0$.}
\bibitem[{not({\natexlab{c}})}]{note}
\bibinfo{note}{We slightly change the notations: ours $\omega_{\pm}$ and
  $\pm\mu_{\pm}$ correspond to $\omega_{1,2}$ and $\mu_{1,2}$ in
  Ref.~\cite{imambekov08}, respectively. We also denote the density of
  particles $n$ and the Fermi wavevector for quasiparticles $q_0$ instead of
  $D$ and $q$ used in Refs.~\cite{korepin93:book,imambekov08}, respectively.}
\end{thebibliography}

\end{document}